\documentclass{bmcart}

\RequirePackage{hyperref}
\usepackage[utf8]{inputenc} 
\usepackage{graphicx}
\usepackage{amsmath}
\usepackage{array}
\newcolumntype{x}[1]{>{\centering\arraybackslash\hspace{0pt}}p{#1}}

\startlocaldefs
\endlocaldefs

\begin{document}

\begin{frontmatter}

\begin{fmbox}
\dochead{Research}


\title{Evaluating the Construct Validity of Text Embeddings with Application to Survey Questions}


\author[
   addressref={aff1},                   
   corref={aff1},                       
   email={q.fang@uu.nl}   
]{\inits{Q}\fnm{Qixiang} \snm{Fang}}
\author[
   addressref={aff2},
   email={d.p.nguyen@uu.nl}
]{\inits{D}\fnm{Dong} \snm{Nguyen}}
\author[
   addressref={aff1, aff3},
   email={d.l.oberski@uu.nl}
]{\inits{DL}\fnm{Daniel L} \snm{Oberski}}


\address[id=aff1]{
  \orgname{Department of Methodology \& Statistics, Utrecht University}, 
  \street{Padualaan 14},                     %
  \city{Utrecht},                              
  \cny{NL}                                    
}
\address[id=aff2]{%
  \orgname{Department of Information \& Computing Sciences, Utrecht University},
  \street{Princetonplein 5},
  \city{Utrecht},
  \cny{NL}
}
\address[id=aff3]{%
  \orgname{Department of Biostatistics and Data Science, Julius Center, University Medical Center Utrecht (UMCU)},
  \street{Universiteitsweg 100},
  \city{Utrecht},
  \cny{NL}
}



\end{fmbox}


\begin{abstractbox}

\begin{abstract} 
Text embedding models from Natural Language Processing can map text data (e.g. words, sentences, documents) to supposedly meaningful numerical representations (a.k.a. text embeddings). While such models are increasingly applied in social science research, one important issue is often not addressed: the extent to which these embeddings are valid representations of constructs relevant for social science research. We therefore propose the use of the classic construct validity framework to evaluate the validity of text embeddings. We show how this framework can be adapted to the opaque and high-dimensional nature of text embeddings, with application to survey questions. We include several popular text embedding methods (e.g. fastText, GloVe, BERT, Sentence-BERT, Universal Sentence Encoder) in our construct validity analyses. 
We find evidence of convergent and discriminant validity in some cases. We also show that embeddings can be used to predict respondent's answers to completely new survey questions.
Furthermore, BERT-based embedding techniques and the Universal Sentence Encoder provide more valid representations of survey questions than do others. Our results thus highlight the necessity to examine the construct validity of text embeddings before deploying them in social research. 
\end{abstract}





\end{abstractbox}
%

\end{frontmatter}



\section{Introduction}
\label{sec_intro}
Text embedding models, which originate from the field of Natural Language Processing (NLP), can map texts (e.g. words, sentences, articles) to supposedly semantically meaningful, numeric vectors (i.e. embeddings) with typically a few hundred or even thousand dimensions (e.g. \cite{mikolov_distributed_2013, reimers_sentence-bert_2019}). 
Intuitively, this means that the embeddings of similar texts (e.g. words like ``big" and ``large") would be closer to one another than those of dissimilar texts (e.g. ``big" and ``paper") in the vector space. 

Such models are often \textit{pretrained} on an enormous amount of text data (e.g. Wikipedia, websites, news) and  made publicly available (e.g. \cite{mikolov_distributed_2013, reimers_sentence-bert_2019, bojanowski2016enriching, devlin_bert_2019})). This allows researchers to obtain off-the-shelf pretrained text embeddings for  downstream applications, without the need to spend many computational resources on training the models from scratch. Researchers can also choose to further train the text embedding models on additional task-specific data (i.e. \textit{fine-tuning}) or domain-specific data (i.e. \textit{continued pre-training}) for better performance.


Because of their capability for meaningful text representation and convenient use, text embedding techniques have become increasingly popular and attracted a growing number of applications in social science. For instance, text embeddings have been employed to encode \textit{the Big-Five personality questionnaire} for personality trait prediction \cite{vu_predicting_2020}, \textit{social media posts} for suicide risk assessment \cite{matero_suicide_2019}, \textit{Tweets and TV captions} for emotion detection \cite{de_bruyne_emotional_2021},  \textit{interview data} for automatic qualitative content analysis \cite{grandeit_using_2020}, and \textit{relevant terms} (e.g. gender pronouns, names, occupations) to quantify societal trends of gender and ethnic stereotypes in the US \cite{garg_word_2018}. 



While such applications show the promising potential of text embeddings for social science research, one question is often not addressed: To what extent do text embedding models provide \textbf{valid representations} for the texts of interest and the intended social constructs? Take the study \cite{vu_predicting_2020} as an example, where Vu et al. converted the questions from the Big-Five questionnaire into text embeddings: Do the embeddings encode relevant information about the questions, such as the underlying personality concept\footnote{We use the term ``concept" and ``construct" loosely interchangeably.} of interest and the formulation? Are the embeddings of questions about personality traits with low empirical correlations (e.g. openness and neuroticism \cite{Linden2010TheGF}) located farther away from each other in the vector space than the embeddings of questions about closely related personality traits (e.g. conscientiousness and agreeableness \cite{Linden2010TheGF})? These are important questions, because only when the text embeddings are valid, can a model make use of relevant information in the embeddings for downstream tasks such as personality trait prediction. In contrast, if the downstream task can rely only on spurious characteristics specific to the data at hand, its performance will likely drop when the model is used in a context that is even just slightly different \cite{Geirhos2020ShortcutLI}. Therefore, valid text embeddings provide the basis for robust downstream applications and further understanding of model behaviour. 


We therefore focus on text embeddings' validity for social constructs. By ``validity", we mean \textit{construct validity}, which concerns the degree to which a construct’s operationalisation in a study matches the construct in social scientific theory \cite{Trochim2015ResearchMT}. 
For instance, a survey question which is intended to detect depression in patients but instead measures anxiety, would not be considered a valid instrument, which would subsequently cast doubt on conclusions regarding depression based on this survey question. 
Applying the idea of construct validity, we can view the human comprehension of a piece of text as the construct of interest, and the corresponding embedding as the operationalisation. 
High construct validity is highly regarded in social science, where measurement and explanation are often the research goals. This is also consistent with the aim for more accurate text representation methods and interpretable models in the NLP community. 

There are several types of construct validity \cite{Trochim2015ResearchMT}. We consider the following four main types \cite{deGroot_1994} that apply to our study: 
\begin{enumerate}
    \item \textbf{Content validity} concerns whether the operationalisation adequately covers all relevant aspects of a construct. For instance, a language test with high content validity should cover all the topics relevant to the mastery of the language (e.g. listening, speaking, reading and writing skills);
    \item \textbf{Convergent validity} concerns whether the operationalisation of a construct is highly correlated with different operationalisations of the same or similar theoretical constructs. For instance, a psychological test on stress should highly correlate with a physiological test on stress (i.e. different measurement method for the same construct) and a psychological test on anxiety (i.e. the same measurement method for a highly related construct);
    \item \textbf{Discriminant validity}, in contrast, concerns whether the operationalisation of a construct is poorly correlated with operationalisations of dissimilar theoretical constructs. For example, two ``agree-disagree" survey questions about completely different topics should not correlate; 
    \item \textbf{Predictive validity} concerns the ability of some operationalisation to predict some future target it should relate to (e.g. college aptitude test scores should predict the test taker's  academic performance in college).\footnote{We are aware of the long-standing discussion over the the terminology for (construct) validity testing \cite{newton_standards_2013}. For instance, the Standards for Educational and Psychological Testing \cite{standards_2014} recommends avoiding the use of phrases like the \textit{types of validity}; instead, one should say \textit{the types of evidence for validity} (e.g. content-related, convergent, or discriminant evidence for validity). Meanwhile, this recommendation has had little impact on the actual research practice in various scientific communities \cite{newton_standards_2013}. In this paper, we intentionally adopt the ``types of validity" terminology for the sake of convenience and its more common usage.}
\end{enumerate}

The goal of our paper is to show how to apply the analysis of construct validity (specifically, content, convergent, discriminant and predictive validity) to text embeddings. However, there are two challenges. First, content validity analysis typically relies on the researcher's ability to directly interpret the measure of interest. With text embeddings, this is difficult because the embedding dimensions have no a priori interpretation. Second, studying the other three types of validity often requires checking the correlation between a single numerical summary of a measure (e.g. the mean of a questionnaire scale) and that of another. A text embedding, however, is a high-dimensional numeric vector to which this approach does not apply. Therefore, an alternative analytic approach to construct validity is needed. 

%

To this end, we propose an adapted framework of construct validity analysis for text embeddings. Specifically, we build on tools from the field of interpretable NLP (e.g. \cite{mikolov_distributed_2013, reimers_sentence-bert_2019, Belinkov2021ProbingCP}) and adapt them to construct validity analysis. We demonstrate this framework on one particular type of texts: \textbf{survey questions}. We focus on survey questions for two reasons. First, survey questions are a popular and important measurement tool for social constructs in social science research. Second, they are usually concise and precise texts. This means that there are likely fewer aspects that text embeddings need to encode, compared to texts that are longer and/or are generated in less controlled settings (e.g. Tweets; conversations). 

Through our analysis, we uncover different degrees and types of validity evidence for different text embedding models when they are applied to survey questions. We thus show that it is necessary to examine the construct validity of such measures before using them for a specific research purpose. Building on the findings, we discuss the potential applications and future directions of text embedding techniques for survey and social science research. Our study also contributes an original, publicly available data set consisting of 5,436 survey questions covering various survey concepts and linguistic properties\footnote{\url{https://github.com/fqixiang/Survey-Embedding-Validity/blob/main/data/Synthetic_Questions.xlsx}}. This data set can be used in future research for studying, for example, text embedding models specific for survey research.  
\vspace{\medskipamount}

Our paper is structured as follows. In Section \ref{sec_background}, we introduce both classic count-based text representation techniques and text embedding techniques. In Section \ref{sec_related}, we review current work on applying text embeddings to survey questions and validity research in NLP. In Section \ref{sec_pretrained}, we outline the specific text embedding methods used in our study and explain how embeddings for survey questions can be computed. Then, we describe how we examine the content validity (Section \ref{sec_content}), convergent \& discriminant validity (Section \ref{sec_con_dis}) and predictive validity (Section \ref{sec_criterion}) of various text embedding approaches for survey questions. We also present the analysis results in each corresponding section. Lastly, we summarise and discuss the findings in Section \ref{sec_conclusion}.

\section{Background: Text Representations}
\label{sec_background}
\subsection{Classic Count-based Text Representation Techniques}
\begin{figure}[htbp]
\includegraphics[width=0.95\textwidth]{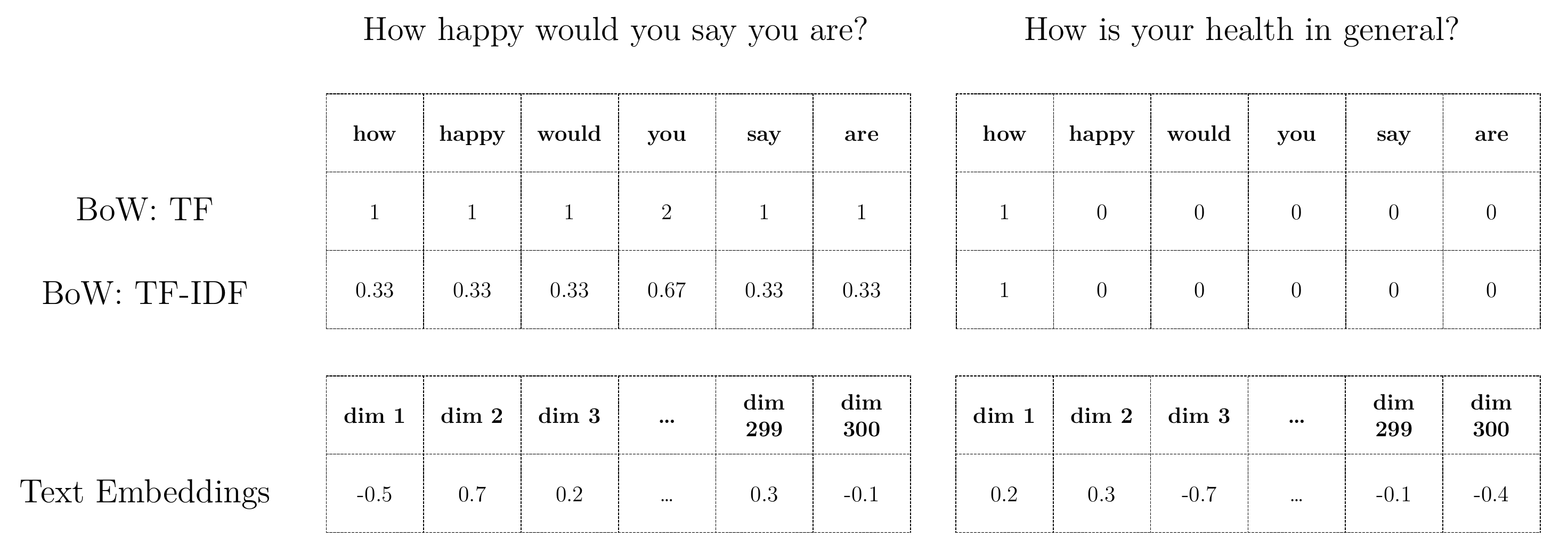}
\caption{\csentence{Different text representation approaches.} For TF (term frequency) and TF-IDF (term frequency-inverse document frequency), the column features represent the vocabulary based on the first survey question.}
\label{fig_text_rep}
  \end{figure}

Prior to the introduction of text embeddings, a popular count-based approach to text representation was bag-of-words (BoW) \cite{Manning2005IntroductionTI}. BoW represents a piece of text by describing the occurrence of words within the text, without taking word order into account (and hence the name). One way to use BoW is to simply count raw term frequency (TF). Take the survey question “How happy would you say you are?” as an example, we can represent this question as a vector [1, 1, 1, 2, 1, 1], with the numbers corresponding to the frequency of the terms “how”, “happy”, “would”, “you”, “say” and “are” (see Figure \ref{fig_text_rep}). 

However, a problem with raw term frequency is that highly frequent words (e.g. “the”, “and”) are not necessarily important words. A different approach, term frequency-inverse document frequency (TF-IDF), mitigates this issue by rescaling words according to how often they appear in all documents (i.e. document frequency). In this way, frequent but often uninformative words like “the” are penalised. Specifically, TF-IDF is calculated as: $\text{tf}_{i,j} \cdot \log(\text{N}/\text{df}_i)$, where $\text{tf}_{i,j}$ refers to the number of occurrences of word $i$ in document $j$, $\text{df}_i$ is the number of documents containing word $i$, and $\text{N}$ is the total number of documents. Note that a document can be a sentence, a paragraph, a book, etc.

Both approaches share many strengths and weaknesses. For example, they are simple and efficient, but at the cost of potentially ignoring relevant information like word relations, grammar and word order.
They also require specifying a priori a list of words to define the features, which is normally based on the vocabulary in the training data. This can lead to the problem that it is not possible to encode out-of-sample, and therefore out-of-dictionary, words. Figure \ref{fig_text_rep} illustrates this using a new survey question: “How is your health in general?”. If we define the feature vocabulary solely based on the first question, then for the new question, all the words except ``how” are missing from both TF and TF-IDF representations, meaning that a large amount of meaningful information from question 2 is lost. 

\subsection{Text Embedding Techniques}
\subsubsection{word2vec and fastText}
One influential family of text embeddings algorithms is word2vec  \cite{mikolov_distributed_2013, Mikolov2013EfficientEO}. Simply put, word2vec is a two-layer neural network model that takes as input a large corpus of text and gives as output a vector space. This vector space has typically several hundred dimensions (e.g. 300), with each unique word in the training corpus being assigned a corresponding continuous vector. Such a vector is also called an embedding. The objective of the algorithm is to predict words from surroundings words (or the other way around). In this way, the final word vectors are positioned in the vector space such that words sharing common contexts in the corpus are located closely to one another. 
Under the Distributional Hypothesis \cite{Wittgenstein1958PhilosophicalI, Harris1954DistributionalS}, which states that words occurring in similar contexts tend to have similar meanings, closely located words in the vector space are expected to be semantically similar. The similarity between two word vectors can be measured by cosine similarity. Mathematically, it is simply the cosine of the angle between two vectors, which can be calculated as the dot product of the two vectors divided by the product of the lengths of the two vectors. Cosine similarity scores are bounded in the interval [-1, 1], where -1 indicates complete lack of similarity while 1 suggests the other extreme. 


Word2vec has been shown to produce text representations that capture syntactic and semantic regularities in language, in such a way that vector-oriented reasoning can be applied to the study of word relationships \cite{Mikolov2013LinguisticRI}. A classic example is that the male/female relationship is automatically learned in the training process, such that a simple, intuitive vector operation like ``King - Man + Woman" would result in a vector very close to that of ``Queen" in the vector space \cite{Mikolov2013LinguisticRI}. Many studies have also made use of this characteristic to study human biases (e.g. gender and racial bias) encoded in texts \cite{garg_word_2018, Caliskan2017SemanticsDA, Rice2019RacialBI, Kumar2020NurseIC}. 

One popular extension of word2vec is fastText \cite{bojanowski2016enriching}, which is trained on subwords in addition to whole words. This allows fastText to estimate word embeddings even for words unknown to the training corpora. fastText was shown to outperform its word2vec predecessors across various benchmarks \cite{Mikolov2018AdvancesIP}.



\subsubsection{GloVe}
GloVe, which stands for Global Vector (for word representations) \cite{Pennington2014GloVeGV}, is another popular text embedding model. 
Similar to word2vec, GloVe also produces word representations that capture syntactic and semantic regularities in language. However, a major difference is that GloVe is trained on a so-called global word-word co-occurrence matrix, where matrix factorisation is used to learn word embeddings of typically 25 to 300 dimensions.

\subsubsection{BERT and Sentence-BERT}
More sophisticated embedding techniques have recently become available. 
A prominent one is BERT, which stands for Bidirectional Encoder Representations from Transformers \cite{devlin_bert_2019}. Like word2vec and fastText, BERT is a type of neural networks trained on a large text corpus in order to learn a good representation of natural language. Notably, BERT produces context-dependent embeddings. That is, while word2vec, fastText and GloVe models produce a fixed embedding for each word, BERT can produce different embeddings for the same word depending on the context (e.g. the neighbouring words). For instance, the word ``bank" can have different meanings (e.g. a financial establishment or land alongside a river) or function (i.e. as a noun or verb), across linguistic contexts.

BERT has achieved state-of-the-art performance on various natural language tasks such as Semantic Textual Similarity, Paraphrase Identification, Question Answering, and Recognising Textual Entailment \cite{devlin_bert_2019}. BERT embeddings have also been shown to encode syntactic and semantic knowledge about the original texts \cite{rogers_primer_2020}. 

Sentence-BERT \cite{reimers_sentence-bert_2019}, an extension of BERT,  differs from the original BERT in that its architecture is optimised for generating semantically meaningful sentence embeddings that can be compared using cosine similarity \cite{reimers_sentence-bert_2019}.


\subsubsection{Universal Sentence Encoder}
Universal Sentence Encoder (USE) is another text embedding model meant for greater-than-word length texts like sentences, phrases and short paragraphs  \cite{Cer2018UniversalSE}. 
USE uses both a Transformer model and a Deep Averaging Network model. The former focuses on achieving high accuracy despite suffering from greater resource consumption and model complexity, while the latter targets efficient inference at the cost of slightly lower accuracy \cite{Cer2018UniversalSE}. It is trained on various language understanding tasks and data sets, with the goal to learn general properties of sentences and thus produce sentence-level embeddings that should work well across various downstream tasks. Pretrained USE embeddings have been shown to outperform word2vec-based pretrained embeddings across different language tasks.  

\subsubsection{Other Techniques}
We are aware of other (near) state-of-the-art text embedding techniques, such as GPT-3 \cite{Brown2020LanguageMA}, ALBERT \cite{Lan2020ALBERT} and XLNet \cite{Yang2019XLNetGA}. Nevertheless, it is not our goal to cover all potential text embedding methods. We limit our analyses to fastText, GloVe, BERT, Sentence-BERT and USE.

\section{Related Work}
\label{sec_related}
\subsection{Research Applications of Text Embeddings to Survey Questions}
We have identified two studies that applied text embedding techniques specifically to survey questions.
Vu et al. \cite{vu_predicting_2020} used BERT to encode participants’ social media posts and the questions from the Big-Five personality questionnaire. By making use of the generated pretrained text embeddings, they were able to moderately improve the prediction of individual-level responses to out-of-sample Big-Five questions, compared to not using any embeddings.
Pellegrini et al. \cite{SonabendW2020IntegratingQM} used the skip-gram embedding algorithm \cite{mikolov_distributed_2013} to represent the questions in 9 different questionnaires about psychiatric symptoms. The embeddings of the questions were then weighted by the numerical responses from psychiatric patients, indicating the severity of specific disease symptoms. In this way, the authors created so-called embeddings profiles unique for every patient. They showed that by applying clustering and classification techniques, such embeddings profiles can be used for effective diagnosis of axis I disorders. 

\subsection{Validity Assessment of Text Embeddings}
To the best of our knowledge, there is no prior work directly applying the framework of construct validity to text embeddings. However, there are two main lines of related work. The first line focuses on what information text embedding models encode, such as syntactic knowledge (e.g. parts of speech, negation, word order; \cite{tenney2018what, Liu2019LinguisticKA}), semantic knowledge (e.g. entity types, relations, numbers; \cite{Tenney2019WhatDY, Ettinger2020}) and world knowledge (e.g. common sense reasoning, visual and perceptual properties of objects; \cite{Forbes2019DoNL, Da2019CrackingTC}). This kind of research is comparable to content validity analysis, despite it not being framed as such in the NLP literature. One limitation with such studies is that the relevant aspects that text embeddings need to encode depend on the specific research question. For instance, measuring personality based on texts data would require that different personality dimensions (e.g. openness, extraversion) to be encoded in the embeddings (assuming that such information is available in the text data), while encoding social media posts for suicide risk assessment would need other types of information like feelings and intentions in the text embeddings. Therefore, for a new research task, it is unlikely sufficient to rely on previous research findings to determine whether a text embedding model can provide an adequate degree of content validity. 

The second line of research focuses on validating embedding-based measurements or predictions by inspecting their correlations with external data sources like human judgments (e.g. \cite{reiter-belz-2009-investigation, reiter-2018-structured, vanAtteveldt2021TheVO}) and surveys (e.g. \cite{garg_word_2018, OConnor2010FromTT, Pasek2019WhosTA}). This procedure is analogous to convergent validity analysis.

We can see that either line of work only focuses on one particular type of construct validity. The resulting validity findings thus provide only a limited view. Our paper, on the other hand, brings together both types of validity under the more general construct validity framework where we also consider two additional types of validity: discriminant and predictive validity. 



\section{Pretrained Embeddings for Survey Questions}
\label{sec_pretrained}
\subsection{Pretrained Text Embedding Models}
In this paper, we investigate whether current text embedding techniques like fastText, GloVe, BERT and USE can produce valid representations for survey questions. We focus on pretrained embedding models (as opposed to fine-tuned models) because of their widespread use. However, our approach to construct validity analysis also applies to fine-tuned text embeddings and other vector representations of texts.

\begin{table}[htbp]
\caption{Overview of Pretrained Text Embedding Models Investigated in this Study}
\begin{tabular}{ccccc}
\hline
Model         & Name                 & Dimension & File Size \\ \hline
fastText      & cc.en.300.bin        & 300       & 2.44 GB \\ 
GloVe         & glove.840B.300d      & 300       & 2.03 GB \\ 
BERT          & BERT-base-uncased    & 768       & 420 MB \\
BERT          & BERT-large-uncased   & 1024      & 1.25 GB \\
Sentence-BERT & All-DistilRoBERTa-V1 & 768       & 292 MB \\
Sentence-BERT & All-MPNet-base-V2    & 768       & 418 MB \\ 
USE           & USE-V4               & 512       & 916 MB \\\hline
\end{tabular}
\label{tab:pretrained}
\end{table}

Table \ref{tab:pretrained} lists the pretrained embedding models we adopt. For fastText, we use the pretrained model developed by \cite{bojanowski2016enriching}. It is trained on Common Crawl with 600B tokens, and produces word embeddings with 300 dimensions for 2M words. For GloVe, we use the model pretrained on Common Crawl with 840B tokens and a vocabulary of 2.2M words. It also outputs word vectors of 300 dimensions. 

As for pre-trained Sentence-BERT models, there are many to choose from, which differ not only in the specific natural language tasks that they have been optimised for, but also in their model architecture. 
We select two pretrained models which have been trained on various data sources (e.g. Reddit, Wikipedia, Yahoo Answers; over 1B pair of sentences) and are thus designed as general purpose models \cite{reimers_sentence-bert_2019}. They are ``All-DistilRoBERTa" and ``All-MPNet-base", where ``DistilRoBERTa" \cite{liu_roberta_2019, Sanh2019DistilBERTAD} and ``MPNet" \cite{song_mpnet_2020} are two different extensions of the original BERT. ``Base" indicates that the embedding dimension is 768, as opposed to ``Large" where the dimension is 1024. Both ``All-DistilRoBERTa" and ``All-MPNet-base" have been shown to have the top average performance across various language tasks. 
For the purpose of comparison, we also include two pretrained models of the original BERT model \cite{devlin_bert_2019}: ``BERT-base-uncased" and ``BERT-large-uncased". ``Uncased" refers to BERT treating upper and lower cases equally. Both models have been trained on Wikipedia (2.5B words) and BookCorpus (800M words). 


As for USE, we use the most recent (i.e. 4th) version of its pretrained model, which outputs a 512 dimensional vector given an input sentence. The sources of training data are Wikipedia, web news,
web question-answer pages, discussion forums and the Stanford Natural Language Inference corpus \cite{Bowman2015ALA}. 

\subsection{Using Pretrained Embedding Models for Sentence-level Embeddings}
\label{sec_sent_emb}
\begin{figure}[htbp]
\includegraphics[width=0.95\textwidth]{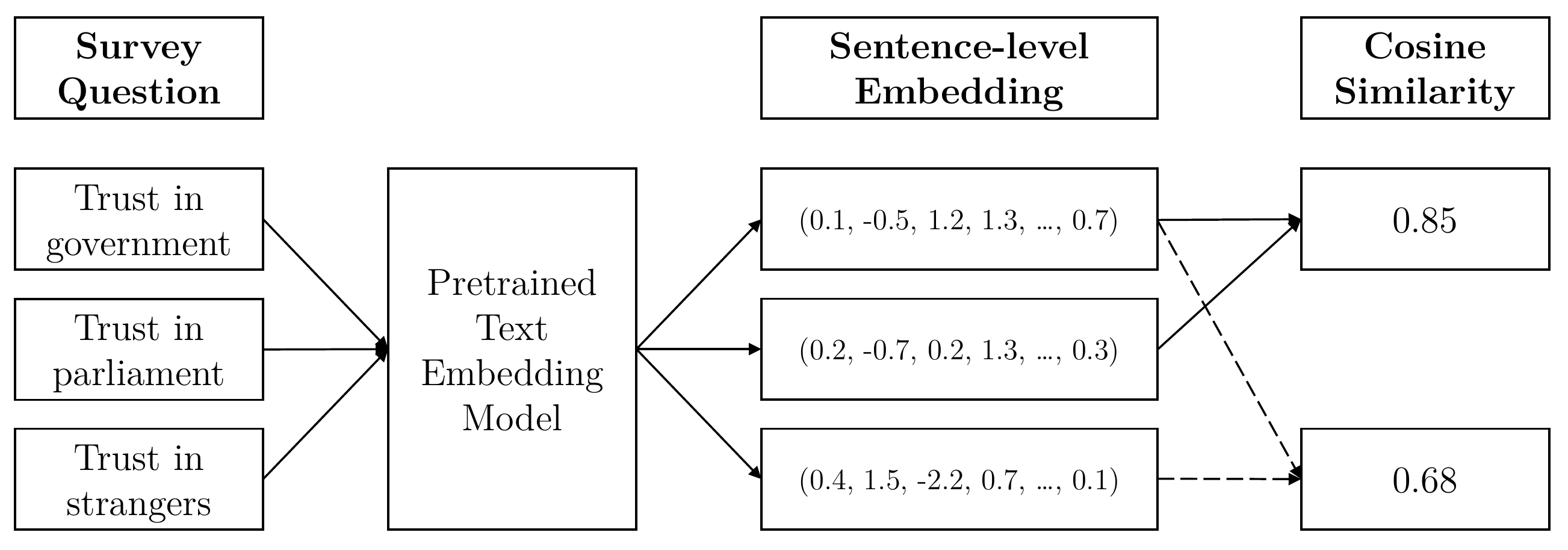}
\caption{\csentence{Illustrating the Use of Pre-trained Text Embedding Models}}
\label{fig_embedding}
  \end{figure}

For texts like survey questions, we need to obtain sentence-level representations (as opposed to word-level) from pretrained text embedding models. Namely, we treat a survey question as one sentence and represent it as a single embedding. Figure \ref{fig_embedding} illustrates this process. For instance, given three survey questions that measure three different concepts (``trust in government", ``trust in parliament" and ``trust in strangers"), we can feed them into a pretrained embedding model and in return obtain three sentence-level embeddings that supposedly represent the original questions. Then, we can calculate cosine similarity scores for these embeddings.


Obtaining sentence-level embeddings is straightforward with Sentence-BERT and USE models, because they have been designed for this specific purpose. In contrast, word2vec and GloVe models only produce word embeddings. When using these models, it is therefore necessary to combine the word embeddings into a sentence-level representation. Although various methods to do so have been proposed, simple averaging across all the word embeddings (e.g. taking the means along each dimension) has been shown to be either outperform other methods \cite{Tawfik2020EvaluatingSR} or approximate the performance of more sophisticated ones \cite{Rckl2018ConcatenatedPW}. Therefore, we use simple averaging to compute sentence-level embeddings for survey questions from fastText and GloVe word embeddings. The resulting embeddings have the same number of dimensions as the word-level embeddings, as we average the word embeddings along each dimension. However, one disadvantage of this approach is that information like word order is absent in the aggregated representation. 

As for the original BERT models, we follow the advice of 
\cite{reimers_sentence-bert_2019, rogers_primer_2020} to average the word embeddings produced at the last layer of BERT to form sentence-level embeddings. This way, the resultant sentence-level representation has the same dimension as that of the word embeddings.



\section{Analysis of Content Validity}
\label{sec_content}
Our analysis of content validity concerns whether text embeddings encode information about all aspects of survey questions relevant to human comprehension. Naturally, not all aspects are equally important, and we also cannot provide an exhaustive list of them. In this paper, we consider four such aspects. 

The first aspect concerns the underlying \textbf{concepts}. According to the typology proposed by \cite{saris_design_2007}, most survey questions can be categorised into one of 21 so-called \textbf{basic concepts}, such as ``feelings", ``cognitive judgement" and ``expectations" (see Appendix A). In addition to the basic concept, a survey question also has a \textbf{concrete concept}, such as ``happiness" (under the basic concept ``feelings") and ``political orientation" (under ``cognitive judgement"). 

Furthermore, survey questions can differ in terms of \textbf{formulation}. Specifically, five types of  formulation often apply in survey research \cite{saris_design_2007}: direct request (DR), imperative-interrogative request (ImIn), interrogative-interrogative request (InIn), declarative-interrogative request (DeIn) and interrogative-declarative request (InDe)\footnote{\cite{saris_design_2007} mentioned one more formulation type: direct instruction, which does not apply to most survey questions concerning subjective basic concepts and is thus not considered in our study.}. See Appendix B for examples of these different formulations. 

Lastly, \textbf{complexity} is another important aspect of survey questions which can affect how respondents understand and answer a survey question \cite{Yan2008FastTA}. A simple proxy for complexity is the length of a survey question \cite{saris_design_2007}. 

Therefore, we investigate whether text embeddings encode information about the following aspects of survey questions: basic concepts, concrete concepts, formulation and length. We refer to them as \textbf{properties} in the remainder of the paper. 

\subsection{Data}
\label{sec:synthetic}
We construct a synthetic data set of survey questions that satisfies two requirements. First, it should cover a wide, (ideally) representative range of (basic and concrete) survey concepts. Second, for every survey question, there should be corresponding survey questions that differ in only the concepts, or only the formulation, or both (similar to the idea of ``minimal pairs" in NLP \cite{Warstadt2020BLiMPAB}). In this way, we have better control over the properties of the survey questions, which will benefit our validity analysis. 

For the first requirement, we focus on covering a wide selection of concepts for \textit{subjective} survey questions. Such questions aim to measure information that only exists in the respondent's mind (e.g. opinions). According to Saris and Gallhofer \cite{saris_design_2007}, subjective survey questions normally fall under one of the following 13 basic concepts: ``evaluation", ``importance", ``feelings", ``cognitive judgment", ``causal relationship", ``similarity", ``preferences", ``norms", ``policies", ``rights", ``action tendencies", ``expectation", and ``beliefs". The questions in our data set therefore cover these 13 basic concepts. 

To satisfy the second requirement, for every subjective concept, we assign three reference concrete concepts. Take the basic concept ``evaluation" as an example: we specify ``the state of health services", ``the quality of higher education" and ``the performance of the government" as the three corresponding reference concrete concepts. 
Next, for every reference concrete concept, we specify one similar concrete concept and one dissimilar concrete concept\footnote{Based on our judgment and experience working in the field of survey research.}. 
Finally, for each concrete concept, we create survey questions that vary in their formulation. \cite{saris_design_2007} provided many templates for each type of formulation. We adopt 19 templates and thus created differently formulated survey questions for each concrete concept. Our final data set contains 5,436 unique survey questions. 

\begin{table}[htbp]
\caption{Example Questions from Our Survey Question Data Set. InDe: interrogative-declarative request. DR: direct request.}
\label{tab:dataset}
  \begin{tabular}{>{\centering\arraybackslash}m{1cm} >{\centering\arraybackslash}m{2cm}>{\centering\arraybackslash}m{1cm}>{\centering\arraybackslash}m{1cm}>{\centering\arraybackslash}m{5.5cm}}
  \hline
ID & Concrete Concept & Similarity & Formulation & Survey Question\\ \hline
1 & state of health services & reference & DR & How good is the state of health services in your country? \\ \hline
2 & state of health services & reference & InDe & Do you agree that the state of health services in your country is good? \\ \hline
3 & state of medical services & high & DR & How good is the state of medical services in your country? \\ \hline
4 & state of medical services & high & InDe & Do you agree that the state of medical services in your country is good? \\ \hline
5 & state of religious services & low & DR & How good is the state of religious services in your country? \\ \hline
6 & state of religious services & low & InDe & Do you agree that the state of religious services in your country is good? \\ \hline
\end{tabular}
\end{table}

Table \ref{tab:dataset} shows six example questions from the data set. They all fall under the basic concept ``evaluation". The main concrete concept here is ``the state of health services", while the corresponding similar and dissimilar concepts are ``the state of medical services" and ``the state of religious services". Each concrete concept has two differently formulated questions in the table: DR (i.e. direct request) and InDe (i.e. interrogative-declarative request).

\subsection{Methods}
\subsubsection{Probing Classifiers}
Text embeddings are high-dimensional and opaque, which makes it difficult to understand what information is encoded in them. We adopt a promising approach in NLP: so-called \textbf{probing classifiers}. The idea is to train a classifier that takes text representations as input and predicts some property of interest (e.g. sentence length). If the classifier performs well, this suggests that the text embedding technique has learned information relevant to the property \cite{Belinkov2021ProbingCP}. 

A recommended practice in choosing a classifier is to select a linear model like (multinomial) logistic regression, because a more complex probing classifier may run the risk that it infers properties not actually present in the text representation \cite{Liu2019LinguisticKA, Hupkes2018VisualisationA, Hewitt2019DesigningAI, Alain2017UnderstandingIL, Maudslay2020ATO}. 
Furthermore, it is recommended to always include baselines for comparison \cite{Belinkov2021ProbingCP}. The better the probing classifier based on some text representation performs relative to the baselines, the more evidence that the probed property is present. Following studies like \cite{Belinkov2017WhatDN, Conneau2018WhatYC, Zhang2018LanguageMT, Tenney2019WhatDY}, we include two baselines: simple majority in the training data and random embeddings. To generate random embeddings for each survey question, we randomly generate from a uniform distribution (-1,1) a unique fixed size embedding for each word in the training data.
Then, we simply average the word embeddings along each dimension to derive sentence-level embeddings for the survey questions. 

\subsubsection{Adapting Probing Classifiers to Survey Questions}
A common problem with probing classifiers is that the good performance of the classifier could simply be due to it making use of other properties present in the embeddings that are correlated with the properties of interest \cite{Belinkov2019AnalysisMI}. For instance, if we want to find out whether text embeddings encode information about basic concepts, our training data should differ only in terms of the probed property (i.e. basic concepts). In other words, for survey questions corresponding to a particular basic concept (e.g. ``feelings"), the distribution of other properties should be similar to that of questions belonging to another basic concept (e.g. ``expectation"). Otherwise, we cannot conclude that the performance of our classifier can be explained by whether the text embeddings encode knowledge about basic concepts.

Unfortunately, with natural language data such as survey questions, it is extremely difficult, if not impossible, to construct a data set where properties like features, length and formulation are completely uncorrelated.  
To mitigate this issue, we construct our training and test sets such that they do not share the same distribution of the correlated properties. In this way, the probing classifier can no longer make use of the correlated properties to achieve good performance on the test sets.

In our data, we see that sentence length is highly correlated with all the other properties. Using chi-square tests of independence, sentence length is statistically significantly related to basic concepts ($\chi^2 = 3636.7, df = 36, p < 0.05$), concrete concepts ($\chi^2 = 4612.1, df = 348, p < 0.05$) and formulation ($\chi^2 = 1252.7, df = 12, p < 0.05$), with multiple testing corrected for. Therefore, when probing those properties, we constrain our training data to contain only survey questions that have different lengths than the ones in the test data. Likewise, when probing sentence length, we make sure that our training and test data do not share the same concepts or formulation. 
Furthermore, when probing basic concepts, because concrete concepts are nested within basic concepts (and hence highly correlated), we make sure that the concrete concepts between the training set and the test set do not overlap. 

Unfortunately, even separating the training and test set in terms of sentence length is not enough for effective probing of concrete concepts. We find that regardless of whether we use random embeddings or the actual text embeddings, the classifier always achieves perfect performance on the test set. The absence of difference in performance prohibits us from concluding whether there is any information about concrete concepts encoded in the text embeddings. This is likely due to the fact that the prediction of concrete concepts may rely solely on the presence of certain words, which is a simple task and can be fully captured by even random embeddings. We therefore decided to increase the difficulty of the probing task for concrete concepts. Specifically, we made the classifier predict for a survey question its similar concrete concept (such as ``the importance of achievement" and ``the importance of success") (which we define in Section \ref{sec:synthetic}), while ensuring that the training set and the test set have not seen the exact same concrete concepts. 

Using the probing approaches above, if we observe any positive difference between the performance of the probing classifier and that of the baseline using random embeddings, we can more confidently attribute it to the relevant survey question property being encoded in the text embeddings (on top of simple word-level information). In this way, we can learn about whether one text embedding model encodes more information about a property than does another model.


\subsection{Results}
\label{results_content_validity}

\begin{table}[htbp]
\caption{Results of Content Validity Analysis: Prediction Accuracy Scores of Probing Classifiers. Note that sentence length is converted into a categorical variable with four levels including ``0-10", ``10-12", ``12-15" and ``15-25"; basic concept, concrete concept and formulation are also categorical with 13, 117 and 5 levels, respectively.}
      \begin{tabular}{ccccc}
        \hline
        & Sentence Length & Basic Concept & Concrete Concept & Formulation\\ \hline
        Simple Majority    & 0.389 & 0.010 & 0.029 & 0.255 \\ 
        Random 300         & 0.102 & 0.198 & 0.440 & 0.742 \\
        Random 768         & 0.148 & 0.198 & 0.509 & 0.694  \\
        Random 1024        & 0.074 & 0.198 & 0.548 & 0.731  \\ 
        TF                & 0.148 & 0.198 & 0.636 & 0.770 \\ 
        TF-IDF             & 0.167 & 0.198 & 0.493 & 0.690 \\ \hline
        fastText           & 0.093 & 0.173 & 0.711 & 0.656\\ 
        GloVe              & 0.194 & 0.192 & 0.908 & 0.642 \\ 
        BERT-base-uncased  & 0.657 & 0.175 & 0.815 & 0.944  \\
        BERT-large-uncased & 0.620 & 0.153 & 0.739 & 0.908  \\
        All-DistilRoBERTa  & 0.407 & 0.198 & 0.916 & 0.776 \\
        All-MPNet-base     & 0.481 & 0.198 & 0.929 & 0.805 \\ 
        USE                & 0.454 & 0.198 & 0.903 & 0.853 \\\hline
      \end{tabular}
\label{tab:resultsContentAnalysis}
\end{table}

Table \ref{tab:resultsContentAnalysis} summarises the performance of the probing classifier (multinomial logistic regression) across fastText and GloVe embeddings, two types of original BERT embeddings, two different Sentence-BERT embeddings and the USE embeddings. The baseline classification accuracy scores are based on simple majority voting, random embeddings of three dimension sizes, TF and TF-IDF vectors.

If the classifier performs better on a particular type of text embeddings than on the random embedding baselines for a survey question property, we can conclude that the corresponding text embeddings of survey questions likely encode information about that property.

For sentence length, we see that the BERT-based and the USE embeddings perform better than all the baselines, which suggests that they likely encode information about sentence length. 
Among them, both original BERT models have better performance than any of the Sentence-BERT models. 


For basic concepts, none of the pretrained text embeddings seems able to beat the performance of the baseline random embeddings, TF and TF-IDF vectors. The fact that all text embeddings (including the random embeddings) have similar performance and perform better than the simple majority baseline suggests that only simple word-level information could be used by the classifier. A possible explanation is that the basic concepts as defined by \cite{saris_design_2007} are too abstract for current embedding techniques to learn.

As for concrete concepts, all types of pretrained text embeddings perform much better than the baselines. This suggests that text embeddings likely encode information about concrete concepts of survey questions. We also see that both Sentence-BERT embeddings show better performance than do the original BERT embeddings. This may be because the Sentence-BERT models have been trained on tasks like Semantic Similarity and Paraphrase Identification, which is arguably similar to identifying sentences with similar or identical underlying concrete concepts. USE and GloVe also have similarly good performance. 

Lastly, we can see that random embeddings themselves can already achieve good prediction on the types of formulation, likely because single words are indicative of formulation. This holds true also for TF and TF-IDF. The random embeddings even outperform fastText and GloVe, despite the margin being relatively small. The original BERT representations, like with sentence length, perform the best again, suggesting that they encode sentence-level information about formulation. Both Sentence-BERT models and USE also perform better than the random baselines, however, only to a much smaller margin. 

To conclude, we find that different text embedding techniques encode somewhat different kinds of information about survey questions and to different degrees. 
If we rank the importance of the properties of survey questions in the order of concepts, formulation and sentence length, then USE seems to demonstrate the highest level of content validity with regard to survey questions on average. The sentence-BERT and original BERT models quickly follow. FastText and GloVe as word embedding techniques do encode some information about survey questions like concrete concepts, but not sentence length or formulation. 

\section{Analysis of Convergent \& Discriminant Validity}
\label{sec_con_dis}
We analyse convergent validity of text embeddings for survey questions as the extent to which the text embeddings of two conceptually similar survey questions are similar to each other. High convergent validity (as is desired) would be indicated by a high degree of similarity between the two text embeddings.
In contrast, discriminant validity concerns the degree to which two conceptually dissimilar survey questions differ in their text embeddings. High discriminant validity (as is desired) is signalled by low similarity between the text embeddings. Convergent and discriminant validity are two sides of the same coin. A measure is only properly defined in relation to other measures when both types of validity are established \cite{Trochim2015ResearchMT}. 

\subsection{Data}
We use the same synthetic data set of survey questions presented in Section \ref{sec_content}. 

\subsection{Methods: Cosine Similarity Analysis}
We take a joint approach to examining convergent and discriminant validity. That is, if text embedding models possess both convergent and discriminant validity, the embeddings of conceptually similar survey questions would be closer to one another while the embeddings of conceptually dissimilar survey questions would be further apart. Two hypotheses naturally follow:

With Hypothesis 1, we expect cosine similarity scores to be higher between the embeddings of \textbf{conceptually similar} survey questions than between those of \textbf{conceptually dissimilar} survey questions, with all other aspects of the survey questions being the same.

With Hypothesis 2, we expect cosine similarity scores to be higher between the embeddings of \textbf{conceptually identical but differently formulated} survey questions than between those of \textbf{conceptually dissimilar but identically formulated} survey questions. 

In the survey question data set we create for the analysis of content validity, we can find many pairs of survey questions that only differ in their concrete concepts and those that differ in their formulation but not in their concrete concepts. 

For Hypothesis 1, we first calculate the cosine similarity between the embedding of a given survey question (i.e. $E_{\text{reference}}$) and the embedding of the corresponding conceptually similar question (i.e. $E_{\text{similar}}$). Then, we calculate the cosine similarity between $E_{\text{reference}}$ and the embedding of the corresponding conceptually dissimilar question (i.e. $E_{\text{dissimilar}}$). In this way, we obtain $\cos{(E_{\text{reference}}, E_{\text{similar}})}$ and $\cos{(E_{\text{reference}}, E_{\text{dissimilar}})}$. We expect the difference between these two scores (i.e. $\cos{(E_{\text{reference}}, E_{\text{similar}})} - \cos{(E_{\text{reference}}, E_{\text{dissimilar}})}$) for a given survey question to be larger than zero. As an example, in Table \ref{tab:dataset}, the two scores of interest are $\cos{(E_{\text{ID1}}, E_{\text{ID3}})}$ and $\cos{(E_{\text{ID1}}, E_{\text{ID5}})}$. Note that the two comparison questions differ from the reference question only in terms of the underlying concrete concepts; all other aspects like the formulation and sentence length are identical. This applies to all the question triads, which allows us to attribute any observed differences in similarity scores to the differences in the concrete concepts.

For Hypothesis 2, we first calculate the cosine similarity between the embedding of a given survey question (i.e. $E_{\text{reference}}$) and the embedding of the corresponding conceptually identical but differently formulated question (i.e. $E_{\text{identical}}$). Then, we calculate the cosine similarity between $E_{\text{reference}}$ and the embedding of the corresponding conceptually dissimilar but identically formulated question (i.e. $E_{\text{dissimilar}}$). In this way, we obtain $\cos{(E_{\text{reference}}, E_{\text{identical}})}$ and $\cos{(E_{\text{reference}}, E_{\text{dissimilar}}})$. We expect the difference between these two scores (i.e. $\cos{(E_{\text{reference}}, E_{\text{identical}})} - \cos{(E_{\text{reference}}, E_{\text{dissimilar}}})$) for a given survey question to be larger than zero. In the exemplar Table \ref{tab:dataset}, the two scores of interest are $\cos{(E_{\text{ID1}}, E_{\text{ID2}})}$ and $\cos{(E_{\text{ID1}}, E_{\text{ID5}})}$. Note that each comparison question differs from the reference question only in terms of one aspect: either concept or formulation. 

\subsection{Results}

\begin{figure}[htbp]
\includegraphics[width=0.95\textwidth]{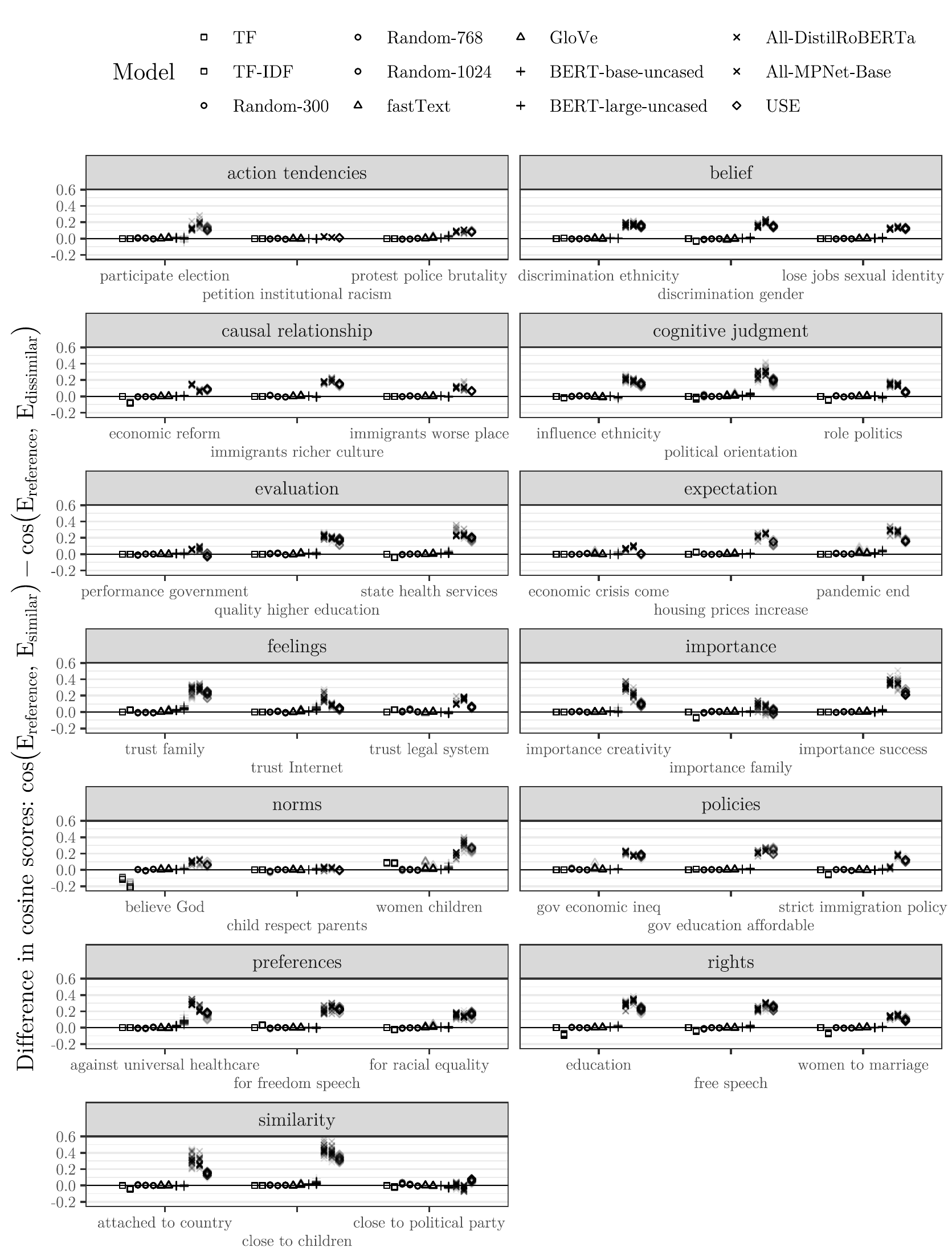}
\caption{\csentence{The distribution of the cosine similarity difference scores for Hypothesis 1 across 13 basic concepts.} The y-axis indicates the size and direction of the differences. The more positive the difference scores are, the more support for convergent and discriminant validity. The x-axis labels are the (abbreviated) names of the main concrete concepts in the question triads.}
\label{fig_h1}
\end{figure}

\begin{figure}[htbp]

\includegraphics[width=0.95\textwidth]{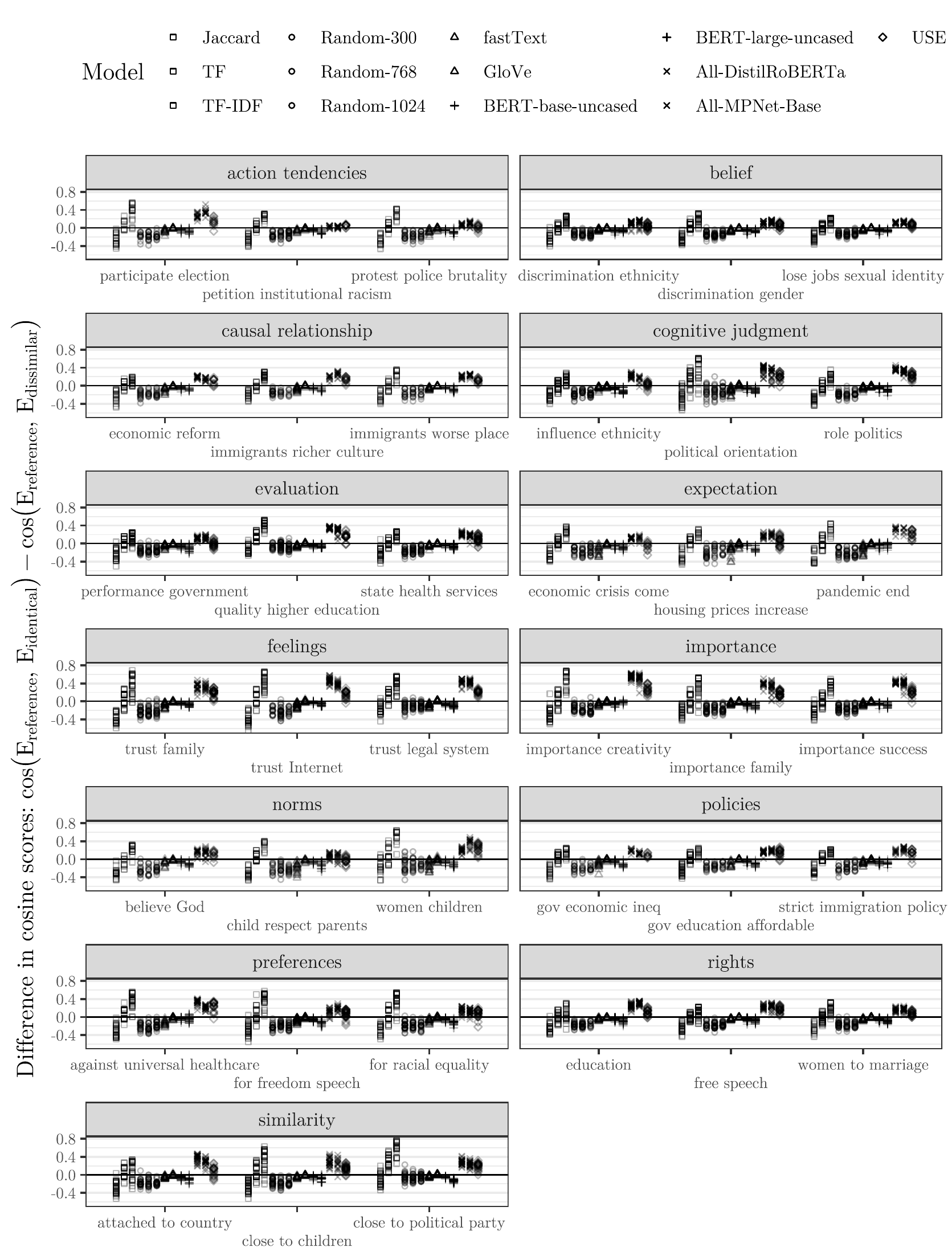}
\caption{\csentence{The distribution of the cosine similarity difference scores for Hypothesis 2 across 13 basic concepts.} The y-axis indicates the size and direction of the differences. The more positive the difference scores are, the more support for convergent and discriminant validity. The x-axis labels are the (abbreviated) names of the main concrete concepts in the question triads.}
\label{fig_h2}
\end{figure}

Figure \ref{fig_h1} shows the distribution of the difference between $\cos{(E_{\text{reference}}, E_{\text{similar}})}$ and $\cos{(E_{\text{reference}}, E_{\text{dissimilar}})}$ scores for Hypothesis 1, across the 13 subjective basic concepts, various baselines and text embedding approaches. The more positive the difference scores are, the more support for convergent and discriminant validity. 

We can see in Figure \ref{fig_h1} that the only models that consistently score above zero are the two Sentence-BERT models ( ``All-DistilRoBERTa" and ``All-MPNet-Base") and USE, with the percentages of positive scores being 98.3\%, 96.8\% and 95.4\%, respectively. This result shows evidence of convergent and discriminant validity for the three models. Only in a small percentage of cases does this observation not hold (e.g. both Sentence-BERT models for the concrete concept ``close to political party" under the basic concept ``similarity"). 
In stark contrast, none of the baselines models (i.e. TF, TF-IDF, random embeddings) show performance comparable to any of the Sentence-BERT and USE models.
To our surprise, this observation holds also for fastText, GloVe and the two original BERT pretrained models, suggesting that these text embeddings approaches lack convergent and discriminant validity.
However, for the original BERT embeddings, one other possible explanation is that cosine similarity might not be a suitable measure, as earlier research suggested \cite{reimers_sentence-bert_2019}. 

Figure \ref{fig_h2} shows the distribution of the difference between  $\cos{(E_{\text{reference}}, E_{\text{identical}})}$ and $\cos{(E_{\text{reference}}, E_{\text{dissimilar}})}$ scores for Hypothesis 2. Note that Jaccard similarity is explicitly included here as an additional baseline of similarity between two survey questions. It is calculated as the ratio of the number of unique overlapping words (i.e. intersection) to the total number of unique words between two survey questions (i.e. union). Naturally, Jaccard similarity scores are bounded in the interval [0, 1]. For Hypothesis 1, because any pair of comparison questions differs in only one word, Jaccard similarity would always be zero and therefore not a useful measure.

Similar to Figure \ref{fig_h1}, the two Sentence-BERT models and USE again score consistently above zero (98.8\%, 97.9\% and 87.2\% of the cases, respectively). Only in a few cases does this observation not hold (e.g. the concrete concept ``petition institutional racism"). We can thus say that for these models, there is reasonable evidence for convergent and discriminant validity. Most of the other approaches (including the baselines, the random embedding models and the original BERT models) score either around or below zero. The only exception is TF-IDF, which in 94.5\% cases scores above zero, suggesting evidence for convergent and discriminant validity. However, this conclusion should be treated with great caution, because when we generate the TF-IDF vectors, we build the vocabulary based on all the survey questions. We follow this approach because in this analysis, it is unclear what the training and testing data should be. In real research applications, TF-IDF may not perform as well due to the difference in the vocabulary between training and test data.

Finally, it is worth noting that the two Sentence-BERT models performed either about equally or better in Hypothesis 2 than in Hypothesis 1, in terms of the percentages of scores above zero. This is somewhat surprising considering that the task in the second hypothesis is supposedly more difficult because the survey questions differ in one extra aspect: formulation. 

Overall, we can conclude that text embeddings of survey questions based on Sentence-BERT and USE demonstrate convergent and discriminant validity. Meanwhile, there is not enough evidence to suggest the same for the other approaches.

\section{Analysis of Predictive Validity}
\label{sec_criterion}
Predictive validity concerns how well a measure of interest predicts some target it should be able to. Here, we define the target as observed individual responses to survey questions. That is, if text embeddings exhibit good predictive validity, they should improve the prediction of a respondent's answer to new survey questions, compared to not using text embeddings. Specifically, we can inspect the correlation between the predicted responses and the actual responses. The higher the correlation (compared to some baseline), the more evidence for predictive validity. 

\subsection{Data: European Social Survey Wave 9}
We use the publicly available European Social Survey (ESS) Wave 9 data \cite{ESS09Data}. The ESS is a research-oriented cross-national survey that is conducted with newly selected, cross-sectional samples every two years since 2001 \cite{ESS09Docu}. 
The survey aims to measure attitudes, beliefs and behaviour patterns of diverse populations in Europe, concerning topics like media and social trust, politics, subjective well-being, human values and immigration.
We focus on the UK sample (N = 2204), because the official language of the UK is English, which is consistent with the language of the data on which our text embedding models are pretrained. 

Out of more than 200 questions that were asked to the participants, we select only the ones which measure subjective concepts and are continuously or ordinally scaled, totalling 94 questions. This choice is consistent with the type of survey questions we examined previously during the analysis of content, convergent and discriminant validity analysis. To harmonise the difference in response scales across the survey questions, we rescale all the responses to be between 0 and 1.

In addition to these 94 survey questions and the individual responses to each of them, we included the following background variables for each participant: region, gender, education, household income, religion, citizenship, birthplace, language, minority status, marital past and marital status. 

\subsection{Methods: Prediction Modelling}
\begin{figure}[htbp]
\includegraphics[width=0.95\textwidth]{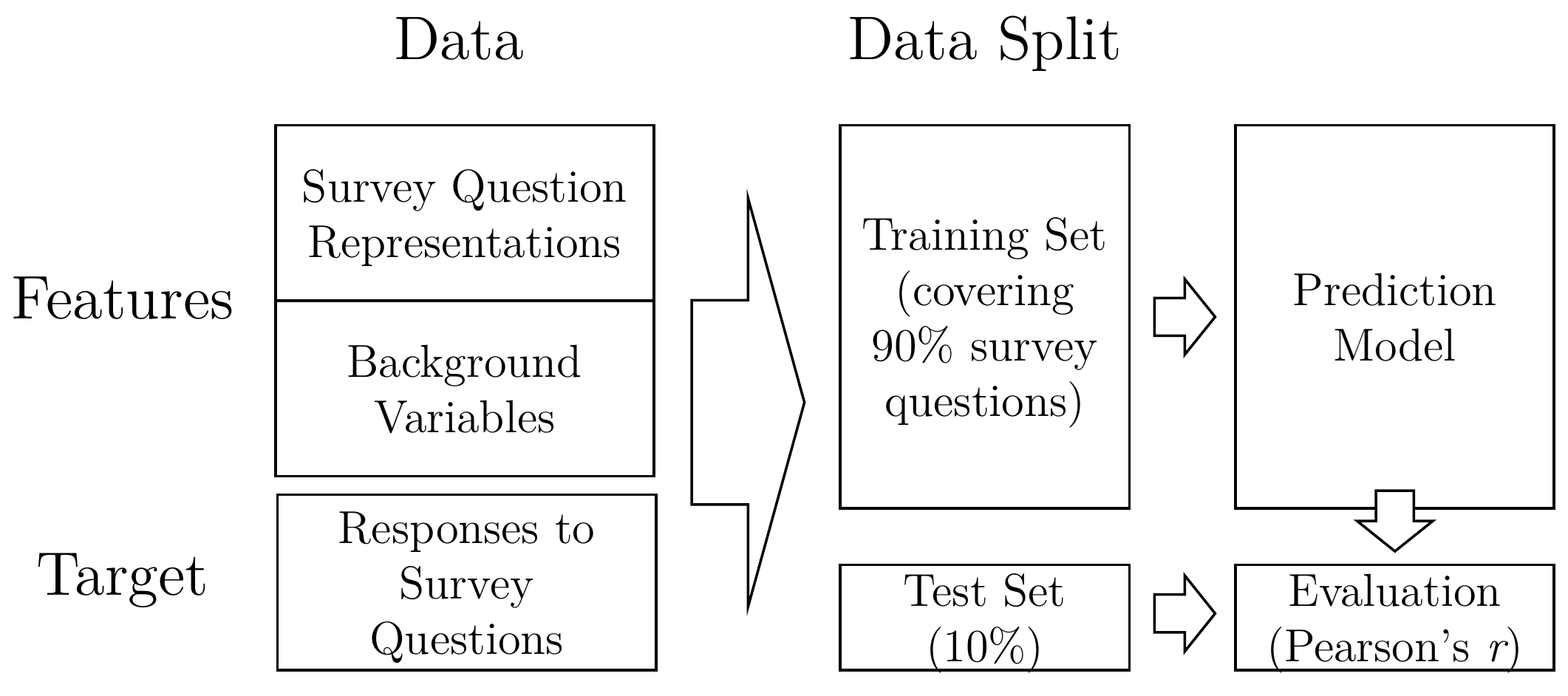}
\caption{\csentence{Prediction Modelling Approach for the Analysis of Predictive Validity.} Note that this figure only applies to one loop in the 10-fold cross-validation process.}
\label{fig_prediction}
\end{figure}

Figure \ref{fig_prediction} illustrates the data set and the prediction modelling process. 
The data set consists of two groups of features: 1) various background variables and 2) text representations of survey questions. The prediction target concerns respondents' answers to survey questions. Each row in the data corresponds to a unique respondent and survey question combination. 

To estimate the average performance of a prediction model, we apply 10-fold cross-validation \cite{Hastie2009}. Namely, we split the data into 10 folds, each containing its own unique survey questions. Then, for every one out of the 10 cross-validation loops, we use 9 folds as the training set (covering 90\% of the survey questions) and keep the remaining 1 fold as the test set. We train the prediction model on the whole training set, also using 10-fold cross-validation but for hyperparameter selection (i.e. fine-tuning). We also ensure that the training-validation splits during the hyperparameter selection are done in such a way that the survey questions in a training partition are different from those in the corresponding validation set. Once we have the fine-tuned model, we apply it to the test set and obtain an evaluation score. Because we apply 10-fold cross-validation to the entire data set, we obtain 10 evaluation scores in total. Finally, we average the 10 scores to arrive at an average performance estimate of the prediction model.

\subsubsection{Text Representation Techniques}
Apart from the various embedding techniques, we include TF, TF-IDF and random embeddings of dimension 300, 768 and 1024 as baseline text representation techniques. Note that with these baseline approaches, we build the feature vocabulary based only on the training data, similar to how we conducted the content validity analysis earlier. That is, new words encountered in the test set would be assigned zero weight.

\subsubsection{Lasso and Random Forest}
We adopt two popular prediction models. The first is Lasso regression \cite{Tibshirani1996RegressionSA}, which differs from OLS regression by including an additional regularisation term in the loss function. The regularisation term has the advantage of reducing model variance (i.e. lower prediction error, at the cost of slightly higher bias). Furthermore, it can zero-out the parameter estimates of those predictors considered by the model to be ``unimportant", thus simplifying the model and easing interpretation.


The second model is Random Forest (RF) \cite{Breiman2001RF}, which constructs multiple regression trees during training time and outputs the average prediction of all the trees. This approach of combining multiple models falls under the so-called ensemble learning technique, which generally provides the benefit of more powerful prediction. In addition, Random Forest automatically considers interaction among the predictors, which Lasso regression falls short of. This may enable Random Forest to learn more fine-grained patterns from data.

\subsubsection{Evaluation Metric}
We use Pearson’s correlation $r$ to evaluate the predictive validity of text embeddings. Specifically, we measure the Pearson’s correlation between the predicted responses to survey questions and the observed responses.
As a \textbf{prediction baseline} (as opposed to TF, TF-IDF and random embeddings, which are baselines for text embedding techniques), we use the average response of each participant in the training data as the prediction for that participant's responses in the test set.

\subsection{Results}

\begin{table}[htbp]
\caption{Results of the Predictive Validity Analysis. $r$ is the average Pearson's correlation between predicted and observed scores. $\Delta\%$ in the parentheses indicates the percentage change in $r$ in comparison to the baseline $r$: 0.187. $95\%$ CI refers to the $95\%$ confidence interval around $r$.}
\begin{tabular}{ccccc}
\hline
                   & Lasso $r$ ($\Delta\%$) & Lasso $95\%$ CI & RF $r$ ($\Delta\%$) & RF $95\%$ CI \\ \hline
TF                 & 0.106 (-43.316)     & [0.102, 0.110] & 0.337 (80.007)  & [0.333,0.341]\\ 
TF-IDF             & 0.092 (-50.802)     & [0.087, 0.096] & 0.323 (72.830)  & [0.319,0.327]\\ \hline
Random 300         & 0.149 (-20.321)     & [0.144, 0.153] & 0.331 (77.066)  & [0.327,0.335]\\
Random 768         & 0.116 (-37.968)     & [0.111, 0.120] & 0.334 (78.614)  & [0.330,0.338]\\
Random 1024        & 0.069 (-63.102)     & [0.065, 0.073] & 0.338 (80.520)  & [0.333,0.342]\\ \hline
fastText           & 0.204 (9.261) & [0.200, 0.209] & 0.356 (90.439)  & [0.352,0.360]\\ 
GloVe              & 0.107 (-42.781)     & [0.103, 0.111] & 0.347 (85.664)  & [0.343,0.351]\\ 
BERT-base-uncased  & 0.195 (4.278)     & [0.191, 0.200] & 0.411 (119.994) & [0.407,0.415]\\
BERT-large-uncased & 0.151 (-19.251)     & [0.147, 0.155] & 0.378 (102.260) & [0.374,0.382]\\
All-DistilRoBERTa  & 0.188 (0.535)     & [0.183, 0.192] & 0.374 (100.228) & [0.370,0.378]\\
All-MPNet-base     & 0.119 (-36.364)     & [0.115, 0.123] & 0.406 (117.135) & [0.402,0.410]\\ 
USE                & 0.186 (-0.535)     & [0.182, 0.191] & 0.386 (106.272) & [0.382,0.390]\\ \hline
      \end{tabular}
\label{tab_criterion}
\end{table}

Table \ref{tab_criterion} summarises the prediction performance of all text representation methods, measured as the average Pearson's correlation $r$ across all 10 folds. $\Delta\%$ in the parentheses indicates the percentage change in $r$ in comparison to the prediction baseline $r$. Higher scores mean that the observed response scores are more correlated with the predicted response scores than with the prediction baseline. The more positive the value of $\Delta\%$, the more evidence for predictive validity. The $95\%$ confidence intervals (CI) around $r$ are also provided. The prediction baseline $r$ is 0.187 with a $95\%$ CI of [0.184, 0.190]. 

We can make two observations.
First, RF consistently performs substantially better than Lasso regression. This is not a surprising result, as we know that RF can learn more complex patterns (like interactions) from data and impose stronger model variance reduction, compared to Lasso regression. Furthermore, because Lasso regression performs worse than or about the same as the prediction baseline, we can infer that the interaction between background variables and the survey question representations is crucial for a good prediction performance.

Second, despite RF faring much better, the exact $r$ scores depend on the text representation methods used. We see that simply using RF with baseline text representation methods (TF, TF-IDF and random embeddings) can already lead to substantial prediction improvement compared to the baseline prediction. All the embeddings approaches achieved  higher $r$ scores than did TF, TF-IDF and random embeddings. This suggests that predictive validity holds (to some extent) for the embedding models, especially the BERT-based models and the USE, which are the highest scoring models.



In summary, we see that text embeddings exhibit some degree of predictive validity. Namely, text embeddings can be used to some extent to predict respondents' answers to completely new survey questions. However, the level of predictive validity that can be demonstrated seems to highly depend on both the specific prediction algorithm and the embedding model used. 

\section{Conclusion and Discussion}
\label{sec_conclusion}
In this paper, we argue that it is important to ensure that text embeddings are valid representations of the original texts and the underlying constructs in social science research. We propose an adapted framework of construct validity analysis for text embeddings and demonstrate it on the case of survey question representation. More concretely, we investigate the content, convergent, discriminant and predictive validity of various popular pre-trained text embedding models, including fastText, GloVe, BERT, Sentence-BERT and USE. Note that despite our focus on pretrained embeddings, our approach to construct validity applies equally to fine-tuned and continued pre-trained embeddings, as well as other vector representations for texts.

For content validity analysis, we propose using probing classifiers from the interpretable NLP literature. Evidence of content validity is indicated by a positive difference in the classifier's performance between using text embeddings and random embeddings. For convergent and discriminant validity analysis, we recommend constructing minimal pairs where the texts only differ on a main property of interest. Then, we can inspect whether the cosine similarity score for a pair of texts is in line with our theoretical expectation. As for predictive validity, we need to define a relevant prediction task whose observed scores can be seen as prediction target. Then, we compute the correlation between the predicted and the observed scores. The higher the correlation, the more evidence for predictive validity. 

When it comes to survey questions, we find that different text embedding techniques demonstrate different degrees of evidence for construct validity. For instance, the USE and the two Sentence-BERT models show the best overall performance across all the construct validity analyses. In contrast, fastText fails to achieve performance comparable to the BERT-based approaches and the USE on all the validity analyses. 
Furthermore, even for the same text embedding model, its performance varies depending on the specific type of construct validity analysis. 
For instance, in our probing experiments to assess content validity, the USE and the two Sentence-BERT models show the best overall content validity. However, the original BERT models perform best in probing the formulation of survey questions. GloVe achieves prediction accuracy comparable to the Sentence-BERT models when probing concrete concepts, but it does poorly on the other probing tasks. 

In light of these findings, we urge researchers to examine and compare the construct validity of different text embedding models before deploying them in any research project. We also hope to see development in text embedding models that can further improve the encoding of abstract information like basic concepts. 

It is also worth noting that our approach to predictive validity can be seen as an example application of using text embedding techniques for survey research. We show that text embedding techniques, when used with the right prediction model, can help to substantially improve the prediction of individual responses to survey questions. This is an exciting result, but we should also realise that the best prediction score ($r$) is still only 0.411, which may be insufficient for production (e.g. in official statistics bureaus). We can likely improve this score by having a larger sample of survey questions.

A limitation of our study is that the synthetic survey question data set and the ESS data do not cover nearly most survey question types (in terms of, for instance, the concrete concepts). This means that our findings concerning the construct validity of the investigated text embedding models may not generalise to other survey questions. Nevertheless, the goal of our paper is not to perform a comprehensive analysis of the construct validity of text embeddings for all survey questions, but to show how to conduct construct validity analysis for text embeddings, with our specific choice of survey questions as an application example.

For future research on the content validity of text embeddings for survey questions, we encourage researchers to probe more complex properties like double-barrelled questions, biases and social desirability. On the application side, we would like to see more research exploring the use of text embedding techniques for, for instance, 1) predicting the quality of new survey questions, 2) detecting problematic survey questions, and 3) generating new survey questions. Lastly, we encourage researchers to apply our construct validity analysis framework to their own research questions.

\section*{Appendix}
\subsection*{Appendix A}

\begin{table}[htbp]
\caption{The 13 Basic Concepts}
\begin{tabular}{c|c}
\hline
Basic Concept       & Example Survey Question \\ \hline
Evaluation          & Is the state of health services in your country is good?  \\ 
Importance          & Is personal success important to you?        \\ 
Feelings            & Do you trust the legal system?       \\
Cognitive judgment  & How large is the role of politics in your life?       \\
Causal relationship & Have immigrants made your country a worse place?        \\
Similarity          & How attached are you to your nationality?       \\ 
Preferences         & Are you for or against universal healthcare?       \\
Norms               & Do you think women should have children?       \\ 
Policies            & Do you agree that the government should reduce income inequality? \\ 
Rights              & Should people have the right to free education in your country? \\
Action tendency     & Will you participate in the election?       \\
Expectation         & Do you expect another economic crisis to come soon?       \\
Belief              & Do people in your country face discrimination because of their gender? \\ \hline
\end{tabular}
\label{tab:13basic}
\end{table}

\subsection*{Appendix B}
\begin{table}[htbp]
\caption{The Five Types of Survey Question Formulation}
\begin{tabular}{c|c}
\hline
Formulation Type       & Example Survey Question\\ \hline
Direct request          & Do you trust the legal system? \\ 
Imperative-interrogative request          & Tell me if you trust the legal system.      \\ 
Interrogative-interrogative request            & Can you tell me if you trust the legal system? \\
Declarative-interrogative request  & I would like to ask you whether you trust the legal system. \\
Interrogative-declarative request  & Do you think that you trust the legal system?    \\ \hline
\end{tabular}
\label{tab:5formulation}
\end{table}



\begin{backmatter}
\section*{Availability of data and materials}
Our code and synthetic data set of survey questions are available at \url{https://github.com/fqixiang/Survey-Embedding-Validity}. The ESS Wave 9 data set is available for download at \url{https://www.europeansocialsurvey.org/download.html?file=ESS9e03_1&y=2018}. The GloVe pretrained embeddings are available at \url{https://nlp.stanford.edu/projects/glove/}. The other pretrained embedding models can be downloaded using the specific Python packages and commands provided in our code.

\section*{Acknowledgements}
The authors thank the following colleagues for their useful feedback (in alphabetical order): Ayoub Baghari, Laura Boeschoten, Anastasia Giachanou, Erik-Jan van Kesteren.

\section*{Competing interests}
The authors declare that they have no competing interests.

\section*{Author's contributions}
DLO proposed the research project; QF designed and performed the research; DN and DLO supervised the research; QF wrote the manuscript. All authors read, proofread and approved the final manuscript.

\section*{Funding}
This work was supported by the Dutch Research Council (NWO) (grant number VI.Vidi.195.152 to D. L. Oberski; grant number VI.Veni.192.130 to D. Nguyen).

\section*{Abbreviations}
NLP, natural language processing; ESS, European Social Survey; BoW, bag-of-words; TF, term frequency;
TF-IDF, term frequency-inverse document frequency; GloVe, Global Vectors for Word Representation; 
USE, Universal Sentence Encoder; BERT, Bidirectional Encoder Representations from Transformers; 
DR, direct request; InDe, interrogative-declarative request; RF, random forest; CI, confidence interval.

\bibliographystyle{bmc-mathphys} 
\bibliography{bmc_article}      

\end{backmatter}
\end{document}